\documentclass[twocolumn,showpacs,pra,aps,superscriptaddress,floatfix]{revtex4-1}

\usepackage{natbib}
\usepackage{amsfonts}
\usepackage{mathrsfs}
\usepackage{amssymb}
\usepackage{bm}
\usepackage{graphicx}
\usepackage[colorlinks=true,linkcolor=blue,citecolor=red,urlcolor=blue]{hyperref}

\usepackage {braket}
\begin{document}

\title{Hermitian and non-Hermitian formulations of the time evolution of quantum decay}

\author{Gast\'on Garc\'{i}a--Calder\'{o}n}
\email{gaston@fisica.unam.mx}
\author{Alejandro M\'{a}ttar}
\affiliation{Instituto de F\'{i}sica, Universidad Nacional Aut\'{o}noma de M\'{e}xico, Apartado Postal 20--364,
M\'{e}xico 01000, Distrito Federal, M\'exico \\ }
\author{Jorge Villavicencio}
\email{villavics@uabc.edu.mx} \affiliation{Facultad de Ciencias,
Universidad Aut\'onoma de Baja California, Apartado Postal 1880,
22800 Ensenada, Baja California, M\'exico}

\date{\today}

\begin{abstract}
This work discusses Hermitian and non-Hermitian formulations for the time evolution of quantum decay, that involve respectively, \textit{continuum wave functions} and \textit{resonant states}, to show that they lead to an identical description for a large class of well behaved potentials. Our approach is based on the analytical properties of the outgoing Green's function to the problem in the complex wave number plane.
\end{abstract}


\maketitle

\section{Introduction}

The theoretical description of quantum decay refers to the time evolution $|\psi_t\rangle=\exp(-iHt/\hbar) |\psi_0\rangle$ of an initial state $|\psi_0\rangle$ in a system characterized by a Hamiltonian $H$. In some  decay problems it is not convenient to separate the Hamiltonian into a part with stationary states and a part which is responsible for the decay, usually  treated to some order of perturbation, but rather to consider the full Hamiltonian $H$ to the system. This is usually the case when the decay originates by tunneling through a classically forbidden region. As is well known, following the work by Khalfin \cite{khalfin58}, if the energy spectra $E$ of the system is bounded by below, \textit{i.e.}, $E \in (0,\infty)$, the exponential decay law cannot hold at long times.  At short times there is also a departure from the exponential decaying behavior which is related, however, to the existence of the energy moments of the Hamiltonian $H$ \cite{khalfin68,muga96,gcrr01}. These both type of behaviors have been confirmed experimentally in recent times \cite{monk06,raizen97}.

The motivation for this work goes back to the early times of quantum mechanics. In 1928, Gamow introduced the notion of resonant state to describe the time evolution of decay of $\alpha$ particles in radioactive nuclei \cite{gamow28}.
In order to describe the above process, Gamow considered solutions to the Schr\" odinger equation which at large distances consist only of purely outgoing waves. This is physically appealing because it yields an outward flux  for the decaying particle outside the interaction region. He realized, however, that the absence of incoming waves in the solution at large distances leads to  complex energy eigenvalues. This was in a way satisfactory because it led to the interpretation of the imaginary part of the energy as the inverse of the lifetime $\tau$ in the exponential decay law $\exp(-t/\tau)$, and thus it provided a theoretical framework for the understanding of  the exponential decay law in quantum mechanics. In fact, his approach constituted one of the first successful applications of quantum mechanics involving  tunneling phenomena and also one of the first theoretical treatments of open quantum systems. However, since the amplitude of resonant states increases exponentially with distance, the usual rules of normalization, orthogonality and completeness do not apply. Nonetheless, over the years in spite of these apparent drawbacks, a consistent theoretical framework involving resonant states evolved over the years \cite{gc10,gc11}. The approach involving \textit{resonant states} represents a non-Hermitian formulation that lies strictly outside the usual Hermitian framework of quantum mechanics. In fact, there is a traditional view that considers the non-Hermitian description of quantum decay involving complex energy eigenvalues as an approximate, phenomenological description that cannot be fundamental because it violates the requirement of unitarity \cite{bender07}. The formulation of decay involving  \textit{continuum wave functions} requires of numerical integration over the wave number for each value of the time \cite{dicus02,muga10}. It turns out, however, that the approaches mentioned above may lead to results for the time evolution of decay that are numerically  indistinguishable from each other \cite{gcmv07}. So the question naturally arises of how these very different formulations of decay lead to the same numerical results.

The aim of this work is to investigate the genesis of the above two formulations, Hermitian and non-Hermitian, for the decay process using the analytical properties of the outgoing Green's function to the problem in the complex $k$ plane.
We intend to answer or at least to throw some light in understanding the question posed  at the end of the above paragraph.

The organization of the paper is as follows. In Sec. \ref{two} some general properties of the time-dependent wave solution involving the outgoing Green's function to the problem are briefly discussed. Subsection \ref{three} refers to the continuum wave solutions and provides a derivation for  time-dependent solution in terms of continuum states.  Subsection \ref{fourth} yields  a discussion on the formalism of resonant states and yields an expression for the time-dependent solution in terms of resonant states. In section \ref{fifth}, we illustrate equivalence between the basis of continuum and resonant states for the $\delta$-shell potential. Finally, Sec. \ref{sixth}  gives the concluding remarks.

\section{Time-dependent solution and the outgoing Green's function}\label{two}

We shall consider a very simple, yet no trivial, description of the time evolution of decay of a particle that is confined initially in a real spherical potential of arbitrary shape in three dimensions. Without loss of generality we restrict the discussion to $s$ waves and it is worth mentioning that the description holds also  on the half-line in one dimension. It is also assumed that the interaction potential $ V(r)$ vanishes after a finite distance, \textit{i.e.} $V(r)=0$ for $r > a$. However,  as discussed below, the results obtained hold also for potentials that go faster than exponentials at very large distances. The units employed here are $\hbar=2m=1$.

Let us therefore write the time-dependent Schr\"odinger equation as
\begin{equation}
\left [i \frac{\partial }{\partial t} - H \right ] \Psi(r,t)=0,
\label{d1}
\end{equation}
where  the Hamiltonian $H= -d^{\,2}/dr^2 + V(r)$.

\begin{figure}
\includegraphics[width=6.5cm]{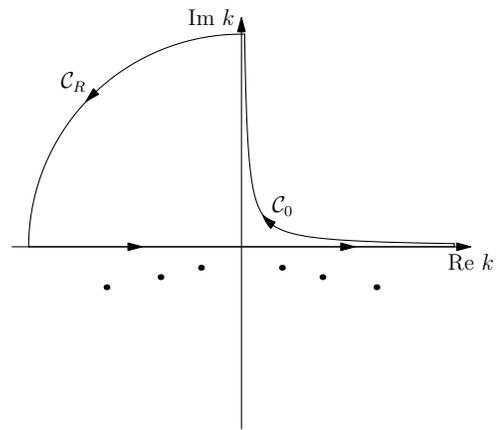}
\caption{Contour $C_0$ and its deformation along the complex k plane used to derive Eq. (\ref{d10}).}
\label{fig1}
\end{figure}

The solution to Eq. (\ref{d1}) may be written as
\begin{equation}
\Psi(r,t)=\int_0^a g(r,r^{\prime};t)\Psi(r',0) \,dr',\quad t >0
\label{d3}
\end{equation}
where $ g(r,r';t)$ stands for the retarded time-dependent Green's function to the problem. This function describes the time evolution of the system  for $t > 0$ and has vanishing value for $t <0$ \cite{newtonchap12}. Clearly, in order to obtain the time-dependent solution one must know $g(r,r';t)$. A convenient form to determine this quantity is by expressing it in terms of the outgoing Green's function to the problem  $G^+(r,r';k)$. For $t > 0$, both quantities are related by using the Laplace transform method \cite{gcp76}
\begin{equation}
g(r,r';t)=\frac{1}{2\pi i} \int_{C_0} G^+(r,r';k) e^{-ik^2t}\,2kdk, \quad t> 0,
\label{d9}
\end{equation}
where $C_0$, before taking the radius of the semicircle $C_R$ up to infinity, corresponds to the contour in the complex k plane shown in Fig. \ref{fig1}.
One may then deform the contour as shown in that figure. The exponential factor in the integrand of (\ref{d9}) guarantees that the contour $C_R$ along the second quadrant of the k plane vanishes as the radius of the semicircle goes to infinity.
As a result $g(r,r';t)$ becomes
\begin{equation}
g(r,r';t)=\frac{i}{2\pi} \int_{-\infty}^{\infty} G^+(r,r';k) e^{-i k^2t}\,2kdk, \quad t> 0.
\label{d10}
\end{equation}
In deriving the above expression we have assumed also for simplicity, since decay refers to a process where the particle tunnels out into the continuum, that the potential does not hold bound states.

Equation (\ref{d10}) is the central quantity to study the dynamics of decay and it is the starting point to analyze the genesis of the descriptions involving \textit{continuum wave functions} and \textit{resonant states}. One sees that it depends on the outgoing Green's function and hence it is convenient to refer first to some of its properties. The  function $G^+(r,r';k)$ obeys the equation
\begin{equation}
[k^2-H]G^+(r,r';k)=\delta(r-r'),
\label{d11}
\end{equation}
and satisfies the outgoing boundary conditions,
\begin{equation}
G^+(0,r';k)=0, \quad G'{^+}(a,r';k)=ik G^+(a,r';k),
\label{d12}
\end{equation}
where the prime denotes here, and thereafter, the derivative with respect to $r$.
The outgoing Green's function  may be written in terms of the so called \emph{regular}, $\phi(k,r)$, and \emph{irregular}, $f_{\pm}(k,r)$, solutions to the Schr\"odinger equation and the Jost function as \cite{newtonchap12}
\begin{equation}
G^+(r,r';k)=-\frac{\phi(k,r_<)f_+(k,r_>)}{J_+(k)},
\label{d13}
\end{equation}
where $r_<$ and $r>$ stand respectively for the smaller and larger of $r$ and $r'$.
We shall discuss briefly some properties of $\phi$,  $f_{\pm}$ and $J_{\pm}$ and then return to discuss  Eq. (\ref{d13}).
The \emph{regular} solution $\phi(k,r)$ satisfies the equation
\begin{equation}
[k^2-H] \phi(k,r)=0,
\label{d14}
\end{equation}
with boundary conditions $\phi(k,0)=0$ and $\phi'(k,0)=1$. The \emph{irregular} solutions $f_{\pm}(k,r)$ satisfy
\begin{equation}
[k^2-H] f_{\pm}(k,r)=0,
\label{d16}
\end{equation}
with boundary conditions
\begin{equation}
f_{\pm}(k,r)=e^{\pm ikr}, \qquad  r > a.
\label{d17}
\end{equation}
Finally, the Jost functions are defined as
\begin{equation}
J_{\pm}(k)=[f_{\pm}\phi'-f'_{\pm}\phi]_{r=a}.
\label{d18}
\end{equation}
Also, $J_{\pm}(k)=f_{\pm}(k,0)$. The \emph{regular} and \emph{irregular} solutions satisfy, respectively, integral equations that are of the \emph{Volterra} type  and hence  may be solved by iteration \cite{newtonchap12}.
This leads to certain conditions on the behavior of the interaction potential $V(r)$ as a function of the distance $r$.
These are that the first and second moments of $V(r)$ must be finite, which imply, respectively, that the potential is less singular near the origin that  $r^{-2}$ and goes faster than $r^{-3}$ as $r \to \infty$. In addition, it is required that
$\int_0^{\infty} dr\,r|V(r)| e^{\eta r} < \infty$, with $\eta$ a real quantity. In particular, if the potential decreases faster than exponential at infinity, $f_+$ and $f_-$ are entire functions of $k$ in the whole complex $k$ plane. This is the case considered in this work, where the  interaction vanishes beyond a distance, but holds also for other type of potentials, as for example,  potentials with Gaussian tails.

Some relevant properties of $\phi$ and $f_{\pm}$ for real values of $k$ are \cite{newtonchap12}: $\phi(k,r)=\phi(-k,r)=\phi^*(k,r)$ and $f_{-}(k,r)=f_{+}^{*}(k,r)$. Notice that the first relationship in the above expressions implies that $\phi(k,r)$ is real and an even function of $k$. The \textit{irregular} solutions  $f_{+}$ and $f_{-}$ are linearly independent functions and it may be shown that $\phi$ may be written as a linear combination of them
\begin{equation}
\phi(k,r)= \frac{1}{2ik}\left [ J_{-}(k)f_{+}(k,r)-J_{+}(k)f_{-}(k,r) \right ],
\label{d26}
\end{equation}
which along region $r>a$, in view of (\ref{d17}), reads
\begin{equation}
\phi(k,r)= \frac{1}{2ik}\left [ J_{-}(k)e^{ikr}-J_{+}(k)e^{-ikr} \right ].
\label{d26a}
\end{equation}

The Jost functions may also be expressed as \cite{newtonchap12}
\begin{equation}
J_{\pm}(k)= 1+ k^{-1}\int_0^{\infty} dr \sin kr \, V(r) f_{\pm}(k,r),
\label{d27}
\end{equation}
and for real values of $k$  one has $J_{-}(k)=J_{+}^{*}(k)$.

A consequence of the above considerations is that the outgoing Green's function $G^+(r,r';k)$ is single valued and analytical in the whole  complex $k$ plane except at an infinite number of poles that correspond to the zeros of the Jost function $J_+(k)$. For potentials that vanish after a distance these poles are in general simple and we shall assume that this is the case here. One finds, in general, a finite number of them may seat on the positive and negative imaginary $k$ axis, corresponding respectively to bound and antibound states, and thatban infinite number are located on the lower half of the $k$ plane where, due to time-reversal considerations, are distributed symmetrically with respect to the imaginary $k$ axis, corresponding, as discussed below, to resonant states. Thus for a pole $\kappa_n=\alpha_n-i\beta_n$ on the fourth quadrant of the $k$ plane, there corresponds a pole $\kappa_{-n}=-\kappa^*_n$ that seats on the third quadrant.

\subsection{Time-dependent solution in terms of continuum wave functions}\label{three}

Here we obtain an expression for the time-dependent solution given by Eq. (\ref{d3}) as an expansion in terms of \textit{continuum wave functions}. The discussion, of course, involves values of $k$ that are real. It starts by noticing that Eq. (\ref{d10}) may be written as
\begin{equation}
g(r,r';t)=\frac{i}{2\pi}\int_{0}^{\infty}[ G^+(r,r';k)-G^+(r,r';-k)] e^{-ik^2t}\,2kdk.
\label{d29}
\end{equation}
Then using Eqs. (\ref{d13}) and (\ref{d26}) one may write the integrand to Eq. (\ref{d29}) as
\begin{equation}
\frac{i}{2\pi}[ G^+(r,r';k)-G^+(r,r';-k)]2k=\psi^{+}(k,r){\psi^{+}}^{*}(k,r'),
\label{d30}
\end{equation}
where the continuum wave functions are defined as
\begin{equation}
\psi^{+}(k,r)=\sqrt{\frac{2}{\pi}}\frac{k\phi(k,r)}{J_{+}(k)}, \,\,
{\psi^{+}}^{*}(k,r')=\sqrt{\frac{2}{\pi}}\frac{k\phi(k,r')}{J^*_{+}(k)}.
\label{d31}
\end{equation}
Substitution of (\ref{d30}) into (\ref{d29}) gives
\begin{equation}
g(r,r';t)=\int_{0}^{\infty}\psi^{+}(k,r){\psi^{+}}^{*}(k,r') e^{-ik^2t}\,dk, \quad t> 0,
\label{d32}
\end{equation}
and substitution of (\ref{d32}) into (\ref{d3}) allows to write the time-dependent solution as
\begin{equation}
\Psi(r,t)=\int_0^{\infty} C(k)\psi^{+}(k,r) e^{-i k^2t}\,dk,
\label{d34}
\end{equation}
where the expansion coefficient $C(k)$ is given by
\begin{equation}
C(k)=\int_0^a {\psi^{+}}^{*}(k,r')\Psi(r',0)\,dr'.
\label{d35}
\end{equation}
Equation (\ref{d34}) yields the time evolution of the time-dependent solution $\Psi(r,t)$ as an expansion in terms of the
\emph{continuum wave functions} to the problem. Notice that by taking the limit as $t \to 0$ in Eq. (\ref{d32}) yields the closure relationship
\begin{equation}
\int_{0}^{\infty}\psi^{+}(k,r){\psi^{+}}^{*}(k,r')\,dk = \delta(r-r'), \quad t> 0,
\label{d32a}
\end{equation}
which shows that the set of \textit{continuum wave functions} is complete.

The \emph{continuum wave functions} are solutions to the Schr\"odiger equation of the problem
\begin{equation}
[k^2-H] \psi^+(k,r)=0,
\label{d36}
\end{equation}
and satisfy the boundary conditions
\begin{eqnarray}
&&\psi^+(k,0)=0\nonumber \\ [.3cm]
&&\psi^+(k,r)=\sqrt{\frac{2}{\pi}}
\frac{i}{2}\left [e^{-ikr}-\textbf{S}(k)e^{ikr} \right ] \quad r>a,
\label{d37}
\end{eqnarray}
where $\textbf{S}(k)$ is the \textbf{S}-matrix of the problem. A comparison of the second term on the right-hand side of (\ref{d37}) and (\ref{d26a}) allows also to write the \textbf{S}-matrix as
\begin{equation}
\textbf{S}(k)=\frac{J_{-}(k)}{J_{+}(k)},
\label{d38}
\end{equation}
and similarly, a comparison between (\ref{d37}) and (\ref{d26}) leads to  Eq. (\ref{d31}), that relates the \emph{continuum wave functions} with the \emph{regular} solutions and the Jost functions.

\subsection{Time-dependent solution in terms of resonant states}\label{fourth}

\textit{Resonant states} are defined as the solutions to the Schr\"odinger equation
\begin{equation}
[\kappa_n^2-H]u_n(r)=0
\label{d46}
\end{equation}
obeying the boundary conditions,
\begin{equation}
u_n(0)=0, \qquad u'_n(a)=i\kappa_nu_n(a).
\label{d53}
\end{equation}
Notice that the second of the above conditions means that for $r >a$,  $u_n(r) = D_n\exp(i\kappa_n r)$, and hence, as first discussed by Gamow \cite{gamow28}, it involves complex energy eigenvalues $\kappa^2_n=E_n=\mathcal{E}_n-i\Gamma_n/2$, with $\kappa_n=\alpha_n-i\beta_n$,
where $\mathcal{E}_n=\alpha^2_n-\beta^2_n$ and $\Gamma_n=4\alpha_n\beta_n$.

The modern approach to resonant states is based on the analytical properties of the outgoing Green's function on the complex $k$ plane. The aim  is to obtain an expansion of $G^+(r,r';k)$ in terms of its poles.

Let us therefore consider the expression \cite{gc76}
\begin{equation}
J=\frac{1}{2\pi i} \int_{C} \frac{G^+(r,r';k')}{k'-k}dk',
\label{d81}
\end{equation}
where $C$ is a large closed contour of radius $L$ in the $k'$ plane about the origin,
which excludes all the poles $\kappa_n$ and the value $k'=k$, namely,  $C=C_R+c_k+\sum_nc_n$. Choosing $C_R$  in the
\textit{clockwise} direction, and $c_k$ and the contours $c_n$ in the \textit{counterclokwise} direction,
it follows using Cauchy's theorem, that $J=0$, and hence one may write
\begin{eqnarray}
&&2\pi iJ=
-\,\int_{C_R} \frac{G^+(k')}{k'-k}dk'+ \nonumber \\ [.3cm]
&& \sum_n \int_{c_n}\frac {G^+(k')}{k'-k}dk'
+\int_{c_k} \frac{G^+(k')}{k'-k}dk'=0.
\label{d82}
\end{eqnarray}
It turns out that as the radius $L \to \infty$, which increases the number of poles inside the contour $C_R$ up to infinity, the function  $G^+(r,r';k)$, that  appears in the  integral over the circle $C_R$, diverges unless both $r$ and $r'$ are smaller than the potential radius $a$, or one of them has the value $a$ and the other remains smaller than $a$ \cite{gcb79,romo80}.  We denote the above conditions, that guarantees convergence of the resonant sum in (\ref{d82}), by the notation $(r,r')^\dagger \leq a$. One may then use the theorem of residues to evaluate the remaining terms  in (\ref{d82}). This requires to know the  residues $\rho_n$ at the poles $\kappa_n$ of the outgoing Green's function. They follow by adapting to the $k$ plane the derivation given in Ref. \cite{gcp76}, namely,
\begin{equation}
\rho_n(r,r')= \frac{u_n(r) u_n(r')}
{2\kappa_n\left \{ \int_0^au_n^2(r)dr + i u_n^2(a)/2\kappa_n \right \} }.
\label{d77}
\end{equation}
which provides the normalization condition for \emph{resonant states}
\begin{equation}
\int_0^au_n^2(r)dr + i \frac{u_n^2(a)}{2\kappa_n }=1.
\label{d78}
\end{equation}
Notice that for bound states, where $\kappa_b=i\gamma_b$, (\ref{d78})  reduces itself to the usual expression. A orthogonality condition for resonant states follows using Green's theorem for solutions $u_n$ and $u_m$ of (\ref{d46}) and its corresponding boundary conditions, to obtain
\begin{equation}
\int_0^au_n(r)u_m(r)dr + i \frac{u_n(a)u_m(a)}{\kappa_n+\kappa_m }=0.
\label{d78a}
\end{equation}
Hence, using (\ref{d77}) in view of (\ref{d78}) gives the purely discrete expansion
\begin{equation}
G^+(r,r';k) = \sum_{n=-\infty}^{\infty} \frac{u_n(r)u(r'_n)}{2\kappa_n(k-\kappa_n)},
\quad  (r,\,r')^{\dagger} \leq \,a.
\label{d85}
\end{equation}
Substitution of  (\ref{d85}) into  (\ref{d11}) leads to the following relationships
\begin{equation}
\frac{1}{2}\sum_{n=-\infty}^{\infty} u_n(r)u_n(r')=\delta (r-r'),\quad (r,\,r')^{\dagger} \leq \,a,
\label{d87}
\end{equation}
which stands for a closure relation, and the sum rule
\begin{equation}
\sum_{n=-\infty}^{\infty}\frac{u_n(r)u_n(r')}{\kappa_n}=0, \qquad (r,\,r')^{\dagger} \leq \,a.
\label{d88}
\end{equation}
Notice that $1/[2\kappa_n(k-\kappa_n)] \equiv 1/(2k[1/(k-\kappa_n) +1/\kappa_n])$,
and hence (\ref{d85}), in view of (\ref{d88}),  may be written also as,
\begin{equation}
G^+(r,r';k) = \frac{1}{2k} \sum_{n=-\infty}^{\infty} \frac {u_n(r)u_n(r')}{k-\kappa_n}, \quad (r,r')^{\dagger} \leq a.
\label{d90}
\end{equation}
Substituting (\ref{d90}) into (\ref{d11}) yields again (\ref{d87}) and the new sum rule
\begin{equation}
\sum_{n=-\infty}^{\infty}u_n(r)u_n(r')\kappa_n=0, \qquad (r,\,r')^{\dagger} \leq \,a.
\label{d88a}
\end{equation}
Substitution of (\ref{d90}) into (\ref{d10}) provides the resonant expansion for the time-dependent Green's function along the internal interaction region, namely,
\begin{equation}
g(r,r';t) = \sum_{-\infty}^{\infty}u_n(r)u_n(r')M(y_n^\circ), \qquad  (r,r')^{\dagger} \leq a,
\label{d91}
\end{equation}
where $M(y_n^\circ)$ stands for the Moshinsky function \cite{moshinsky52}
\begin{equation}
M(y^{\circ}_n)=\frac{i}{2\pi}\int_{-\infty}^{\infty}\frac {e^{-ik^2t}}{k-\kappa_n}dk=
\frac{1}{2} w(iy^{\circ}_n),
\label{d92}
\end{equation}
with $y_{n}^{\circ}=-e^{-i\pi /4} \kappa_n t^{1/2}$. The function $w(iy^{\circ}_n)$ stands for the Faddeyeva function \cite{abramowitzchap7} for which well known computing algorithms have been developed \cite{poppe90}.

One may also derive an expression for $G^+(r,r';k)$ along the external interaction region $r \geq a$ by noticing that for $r' <a$ and $r\geq a$, $G^+(r,r';k)=G^+(a,r';k)e^{ik(r-a)}$. Substitution of this expression into (\ref{d10}) and expanding $G^+(a,r';k)$, using (\ref{d90}), gives
\begin{equation}
g(r,r\,';t)=\sum_{n=-\infty}^{\infty} u_n(r')u_n(a)M(y_n),\quad r'<a,\, r \geq a,
\label{d107f}
\end{equation}
where $M(y_n)$ stands for the Moshinsky function
\begin{equation}
M(y_n)=\frac {i}{2\pi} \int_{-\infty}^{\infty} \frac{e^{ik(r-a)}e^{-i k^2t}}{k-\kappa_n} dk
\label{d107g}
\end{equation}
with $y_n=e^{-i\pi /4} (1/4 t)^{1/2}[(r-a)-2 \kappa_n t]$.

Substitution of (\ref{d91}) and  (\ref{d107g}) into (\ref{d3}) allows to write the resonant expansion for the time-dependent solution
as
\begin{equation}
\Psi(r,t)= \sum_{-\infty}^{\infty}\left \{\begin{array}{cc}
C_nu_n(r)M(y^{\circ}_n), & r\leq a \\[.4cm]
C_nu_n(a)M(y_n), & r \geq a;
\end{array}
\label{s1}
\right.
\end{equation}
where
\begin{equation}
C_n=\int_0^a  u_n(r) \Psi(r,0) dr.
\label{s2}
\end{equation}
Notice in (\ref{s1}) that at $r=a$, $y_n=y^{\circ}_n$.

It is worth mentioning that using the properties of the Faddeyeva function \cite{abramowitzchap7}, one may derive an explicit analytical expression of the time-dependent function along the internal region of the potential at long times \cite{gc10}
\begin{eqnarray}
\Psi(r,t) \approx \sum_{n=1}^{\infty} C_nu_n(r)e^{-i\mathcal{E}_nt}e^{-\Gamma_nt/2} -  \nonumber \\ [.3cm]
i \eta\,{\rm Im} \left\{\sum_{n=1}^{\infty}\frac{C_nu_n(r)}{k_n^3} \right\} \,\frac{1}{t^{\,3/2}};\,\,r \leq a,
\label{3b}
\end{eqnarray}
where $\eta=1/(4\pi)^{1/2}$. Notice that the sum rule (\ref{d90}) has canceled out exactly the leading $t^{-1/2}$ term in the asymptotic long-time behavior of the Faddeyeva function \cite{abramowitzchap7,gc10}.

The time-dependent decaying solution given by the first term in (\ref{s1}) satisfies the \textit{continuity equation}
$i(\partial / \partial t)|\Psi(r,t)|^2+ (\partial/\partial r) J(r,t) =0$, with $J(r,t)$ the current density, which by integration along the internal interaction region yields the relationship,
\begin{equation}
\Gamma_n= 2\alpha_n\frac{|u_n(a)|^2}{\int_0^a|u_n(r)|^2\,dr}.
\label{3ac}
\end{equation}
In deriving the above expression it is required the derivative of the Faddeyeva function \cite{abramowitzchap7} and  Eqs. (\ref{d78a}) and  (\ref{d88a}).
\begin{figure}[!tbp]
\includegraphics[width=3.3in]{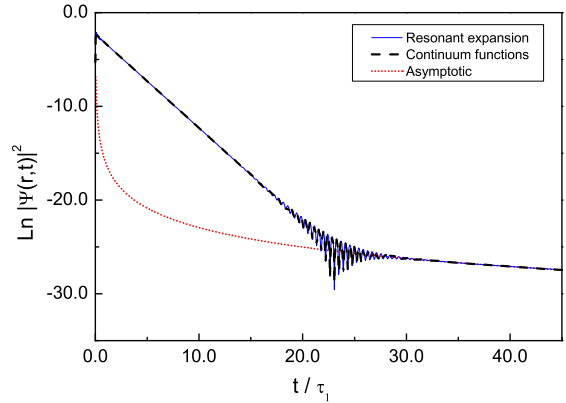}
\caption{Behavior of the probability density in lifetime units at the potential boundary $r=a$ for the $\delta$-shell potential with parameters as indicated in the text.}
\label{fig2}
\end{figure}
\begin{figure}[!tbp]
\includegraphics[width=3.3in]{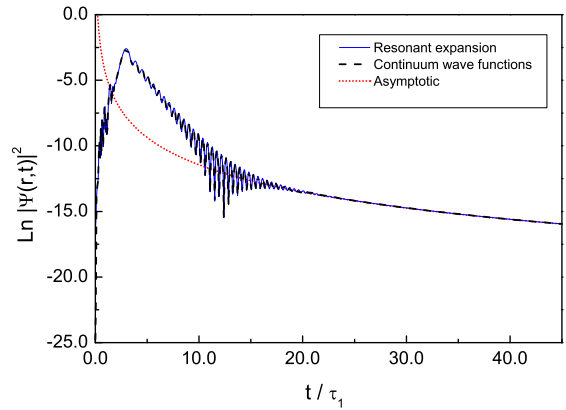}
\caption{Behavior of the probability density in lifetime units at the potential boundary $r=25a$ for the $\delta$-shell potential with  parameters as indicated in the text.}
\label{fig3}
\end{figure}
\section{Model}\label{fifth}

In order to illustrate the Hermitian and non-Hermitian formulations discussed above, we consider a $\delta$-shell potential,
of radius $a$ and intensity $\lambda$ for $s$ waves, namely,
\begin{equation}
V(r)=\lambda \delta(r-a).
\label{5b}
\end{equation}
This model was initially considered by Winter \cite{winter61}, and since then, by many authors. The reason being that its mathematical simplicity does not avoid that it describes correctly the main physical features of the time evolution of decay. As initial state we choose the infinite box state,
\begin{equation}
\Psi(r,0)= \left (\frac{2}{a} \right )^{1/2} \sin \left (\frac{\pi r}{a}\right ).
\label{5bb}
\end{equation}
It follows then, using (\ref{d37})  that the \textit{continuum wave functions}, corresponding to the Hermitian formulation, read
\begin{eqnarray}
\psi^+(k,r) &=& \sqrt{\frac{2}{\pi}} \left \{
\begin{array}{cc}
\sin(kr)/J_+(k),\;\; r \leq a \\[.35cm]
(i/2)\left[e^{-ikr}-  \textbf{S}(k)e^{ikr}\right] & ,\;\; r \geq a,
\end{array}
\right.
\label{11c}
\end{eqnarray}
where the \textbf{S}-matrix is given by (\ref{d38}).  The Jost function $J_+(k)$ of the problem may be obtained immediately using (\ref{d27}), namely, $J_+(k) =2ik+\lambda(\exp(2ika-1)$, and we recall that $J_-(k)=J_+^*(k)$. One then may obtain expressions for the expansion coefficients $C(k)$,  given by (\ref{d35}), and then, by using (\ref{11c}), either along the internal or external regions of the potential,  of  the time-dependent solution $\Psi(r,t)$ given by (\ref{d34}). To obtain the time-dependent solution requires of numerical integration along a span of values of $k$ for each value of time $t$.

On the other hand,  for the non-Hermitian formulation, the resonant states of the problem read,
\begin{equation}
u_n(r)=\left\{
\begin{array}{cc}
A_n \,\sin (\kappa_n r), & r\leq a \\[.3cm]
B_n \,e^{i\kappa_n r}, & r \geq a.
\end{array}
\right.
\label{5c}
\end{equation}
From the continuity of the above solutions and the discontinuity of its derivatives with respect to $r$ (due to the $\delta$-function interaction) at the boundary value $r=a$, it follows that the $\kappa_n$'s may be obtained by solving the equation,
$2i\kappa_n + \lambda ( e^{2i\kappa_n a}-1)=0$, which corresponds to the zeros of the Jost function $J_+(k)$.
For $\lambda \gg 1$ one may write the approximate analytical solutions  to the above equation as,
$\kappa_n \approx (\pi/a) (1-1/\lambda a) -i \,(1/a)(\pi/\lambda a)^2$.
Using the  above expression for $\kappa_n$ as the initial value in the Newton-Rapshon method, \textit{i.e.}, $\kappa_n^{r+1}= \kappa_n^r -
F(\kappa_n^r)/\dot{F}(\kappa_n^r)$, with $\dot{F}=[dF/dk]_{k=\kappa_n}$  yield the solutions $\kappa_n$ with the desired degree of approximation according to the number of iterations.

The normalization coefficients of resonant states may be evaluated by substitution of Eq. (\ref{5c}), for $r \leq a$, into Eq.
(\ref{d78}), to obtain $A_n=[2\lambda/ (\lambda a+ \exp(-2i\kappa_n a))]^{1/2}$. Similarly, using Eqs. (\ref{5bb}), the first expression in (\ref{5c}) and the above expression, yields analytical expressions for the coefficient $C_n$.
Hence, for given values of the potential parameters, $\lambda$ and $a$, one may then calculate the set of complex poles $\{\kappa_n\}$,  resonant states $\{u_n(r)\}$ and coefficients $\{C_n\}$,  to evaluate the time dependent solutions (\ref{s1}). Notice that all these quantities are calculated only once,  which saves considerable computation time.

Figures \ref{fig2} and \ref{fig3}  provide  plots of the $\ln |\psi(r,t)$ as a function of time in lifetime units, $\tau=1/\Gamma_1$,  for  parameters of the $\delta$-shell potential, $\lambda=12.0$ and $a=1$, respectively, for $r/a=1$ and $r/a=25$. One observes in both cases that the Hermitian and non-Hermitian formulations are indistinguishable from each other. Also plotted, in both figures, is  the asymptotic long-time contribution that goes as $t^{-3}$.

\section{Concluding remarks}\label{sixth}

It is worth stressing that the exact Hermitian and non-Hermitian formulations for the description of decay discussed here, corresponding to coherent (elastic) processes,  have as common origin the analytical properties of the outgoing Green's function to the problem and that its equivalence holds for potentials having finite first and second moments and tails that go faster than exponential at long distances or that vanish after a distance. This might allow to consider artificial quantum systems as  ultracold atoms \cite{raizen97} and resonant tunneling structures  \cite{sakaki87}. Since decay by emission of particles corresponds to an open quantum system, it is not surprising that the notion of unitarity usually employed for closed systems does not apply. In this context,  Eq. (\ref{3ac}) is very relevant because it implies flux conservation as time evolves. Finally, it is also worth emphasizing that the \textit{resonant} formalism provides expressions, as (\ref{3b}), which exhibits explicitly the exponential and nonexponential contributions to decay in contrast with the `black box' description (\ref{d34}) involving \textit{continuum wave functions}.

%
%\bibliography{refs}
%

%merlin.mbs apsrev4-1.bst 2010-07-25 4.21a (PWD, AO, DPC) hacked
%Control: key (0)
%Control: author (8) initials jnrlst
%Control: editor formatted (1) identically to author
%Control: production of article title (-1) disabled
%Control: page (0) single
%Control: year (1) truncated
%Control: production of eprint (0) enabled
%

\end{document}